\documentclass[multphys,vecphys]{svmult}

\usepackage{makeidx}         
\usepackage{graphicx}        
\usepackage{multicol}        
\usepackage[bottom]{footmisc}

\begin{document}

\title*{The lack of binaries among hot horizontal branch stars:
M80 and NGC5986}
\titlerunning{Binaries in hot HB stars: M80 and NGC5986}
\author{C. Moni Bidin\inst{1}\and
S. Moehler\inst{2}\and G. Piotto\inst{3}\and Y. Momany\inst{3}\and
A. Recio-Blanco\inst{4}\and R.A. M\'endez\inst{1}}
\authorrunning{Moni Bidin et al.}
\institute{Departamento de Astronom\'{i}a, Universidad de Chile,
Chile
\and European Southern Observatory,
Garching,
Germany
\and Dipartimento di Astronomia, Universit\`{a} di Padova,
Italy
\and Observatoire de la C\^{o}te d'Azur,
Dpt. Cassiop\'ee,
France
}

\maketitle

\section{Introduction}
\label{sec:1}
Extreme horizontal branch (EHB) stars play an important role in extragalactic astronomy, since they have been individuated
as possibly being responsible for the UV upturn in elliptical galaxies and in the bulges of spiral galaxies, that
has been proposed as an independent age indicator for this type of galaxies.
In recent years the ``binary scenario'', in which EHB stars formation is related to dynamical interactions
inside binary systems, has been
proposed as the
main channel for their formation. In fact \cite{Maxted} indicated
that 69$\pm$9\% of field EHB stars should be close binary systems with short periods P$\leq$10 days.
Nevertheless, more
recently \cite{Napiwotzki} found a noticeably lower binary fraction (40-45\%), and
\cite{MoniBidin} found no evidence of binarity among 18 EHB stars in globular cluster NGC6752. They estimated that within
a 95\% confidence level the close binary fraction in EHB of this cluster should be lower than 20\%.

Here we present preliminary results of the extension of the previous survey.

\section{Results}
\label{sec:2}

\begin{figure}
\sidecaption[t]
\centering
\includegraphics[height=8cm]{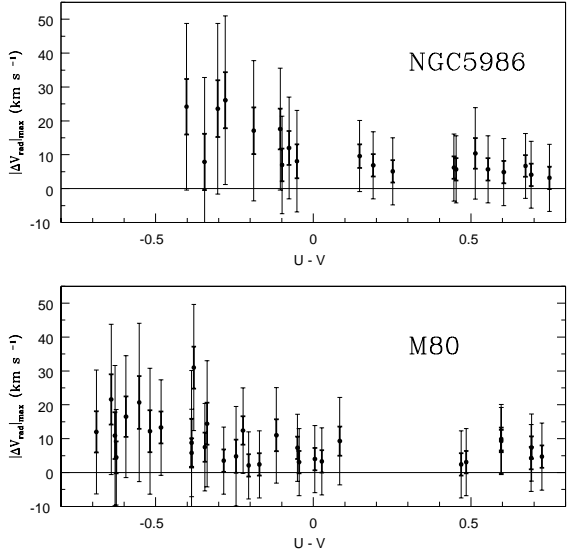}
\caption{Maximum RV variation observed for each target in 4 nights of
observations, as a function of color U-V.
Photometric data are from \cite{Momany2003} and \cite{Momany2004}.
The thick errorbar indicates the 1$\sigma$ interval, the thin one the 3$\sigma$.
Each night we collected up to two spectra per star but observations on M80 were
undersampled due to string wind from north.}
\label{fig1}
\end{figure}

Observations, data reduction, radial velocity (RV) measurements and
error analysis were performed as in \cite{MoniBidin}. Systematic errors must still be rigorously measured,
but the corrections applied here should be within 2-3 km/s from the true value.
Our results are plotted in Figure \ref{fig1}.

The most prominent result is that again we fail to detect the high RV variations observed among many field EHB stars. We
conclude that there is no {\it clear} evidence of binarity in the samples, although we individuate a possible exception in
M80. This star shows a modest (31 km/s) but statistically significant (nearly 5$\sigma$ from zero) variation,
and we consider it an interesting candidate, although the variation is not high enough to rule out the possibility
that it is due to some distortion induced by noise.
In NGC5986 we find one variation slightly higher than $3\sigma$ (26.1$\pm$8.3 km/s, i.e. 3.1$\sigma$),
but it is not trustworthy due to the low S/N of the spectra.
Such a variation is statistically reasonable among the great number of our measurements,
it is hard to consider it significant. No further conclusion can be drawn at the moment.

In our sample we analyzed 11 EHB stars in M80. Our observations fix the best estimate for
the close binary fraction $f$=15\%.
Unfortunately it is not a strong constraint because of the poor temporal sampling, that lowered the
sensitivity of the survey. Statistically considering our observations and results,
the probability that $f\geq$48\% is lower than 10\%,
enough to rule out the high
$f$ observed by \cite{Maxted} but not the intermediate one found by \cite{Napiwotzki}.

In NGC5986 we observed 5 EHB stars. The sample is too much small to attempt any statistical consideration,
but the result
again points out a lack of close binaries, even in the presence of one candidate,
because in this cluster our survey reaches a high detection probability
(80\% on average).

We can then conclude that in both clusters the general lack of close binaries among EHB stars is confirmed.

\end{document}